\documentstyle[12pt,psfig]{article}

\def\baselinestretch{1.2}
\begin{document}

\begin{titlepage}

\begin{flushright}
UMN-D-01-3 \\
hep-th/0011106
\end{flushright}
\vfil\vfil

\begin{center}

{\Large {\bf Renormalizing DLCQ Using Supersymmetry }}

\vfil
I.~Filippov,$^a$  S.~Pinsky,$^a$ and J.R. Hiller$^b$  \\
\vfil
$^a$Department of Physics \\ Ohio State University, Columbus, OH~~43210\\
\vfil
$^b$Department of Physics\\ University of Minnesota-Duluth, Duluth, MN~~55812\\
\vfil
\end{center}

\begin{abstract}
Recent string theory developments suggest the necessity to understand
supersymmetric gauge theories non-perturbatively, in various dimensions. In this
work we show that there is a standard Hamiltonian formulation that generates a
finite and supersymmetric result at every order of the approximation
scheme known as discrete light-cone quantization (DLCQ). 
We present this renormalized DLCQ Hamiltonian and find that it has two
novel features: it automatically chooses the 't Hooft prescription for
renormalizing the singularities, and it introduces irrelevant operators that
serve to preserve the supersymmetry and improve the convergence.  We solve for
the bound states and the wave functions with and without the irrelevant operators
and verify that with the irrelevant operator the  exact large-$N_c$ 
supersymmetric DLCQ (SDLCQ) results are reproduced. 
With the irrelevant operator removed, we show that
the bound-state mass and wave functions appear to be converging to the SDLCQ
results but very slowly. This is a first step in extending the advantages of
SDLCQ to non-supersymmetric theories. 
\end{abstract}

\vfil\vfil\vfil
\begin{flushleft}
March 2001
\end{flushleft}

\end{titlepage}
\renewcommand{\baselinestretch}{1.05}  

\section{Introduction}

Solving for the non-perturbative properties of  quantum field theories --
such as QCD -- is a very difficult problem. In order to gain some
insights, however, a number of lower dimensional models have been
investigated in the large-$N_c$ (planar)  approximation using discrete 
light-cone quantization (DLCQ), with a plethora of examples appearing over
the years (for a review see \cite{bpp98}). Going beyond the simplest
models that have either just fermions or just bosons one encounters
significant problems.

Recall that in the DLCQ approach, the Schrodinger equation for the field
theory is formulated as an infinite set of integral equations for the
wave functions and masses of the bound states of the theory. This infinite
set of integral equations is then truncated and solved numerically.
Problems arise because these integral equations have a number of
singularities that need to be regularized and renormalized before any
numerical solution can be attempted.  In the simplest models one can
follow 't Hooft \cite{tho} and use the principal value prescription to
regulate and effectively renormalize these divergences. There are other
prescriptions besides the 't Hooft prescription, and these 
prescriptions \cite{wu,mal} lead to different results. 
It has only recently been understood
that these other prescriptions produce an incomplete solution \cite{bas00} and
that when a complete set of topological components are included they reproduce
the 't Hooft prescription. In fact, all the singularities that are encountered in
the (1+1)-dimensional formulation can be handled by careful analytical calculations
and the principal value prescription. We will review these calculations to
highlight the difficulties and  ambiguities.

A number of recent string theory developments
have sharpened the need to understand supersymmetric Yang-Mills 
non-perturbatively in
various dimensions, since they play a crucial role in describing 
D-brane dynamics,
the Maldacena conjecture \cite{mal97,ahlp99,hlpt00} and, ultimately, in 
formulating
M(atrix) Theory \cite{bfss97}.  This makes it imperative to develop a 
non-perturbative method to
solve such theories where fermions and bosons are treated on an equal 
footing. The
importance of supersymmetry in string theory and M-theory is the core of the
recent supersymmetric formulation of DLCQ which we
call SDLCQ \cite{mss95}. The principle is to
construct a sequence of approximations to the field theory that converge to the
continuum theory and that remain supersymmetric at
every order of the approximation.

In 1+1 dimensions it is well known that even ${\cal N}=1$ supersymmetric theories
are super renormalizable. Therefore a formulation which does not break
the symmetry will be totally finite, requiring no
regularization or renormalization. In recent years we have solved many such
theories and successfully extended these ideas to higher
dimensions{ \cite{alp98,alp98a,alp99,alp99a,alp99b,alppt98,alpt98,appt98,lup99}.
In this work we would like to return to the original Hamiltonian formulation of
DLCQ and ask what is the regularization and renormalization that exactly
reproduces the SDLCQ formula. The existence of such a formulation has never
really been addressed except in a very simple model \cite{alp99b}. In 
fact, none of the
fully supersymmetric theories that we have solved over the last few years have
been solved using the standard Hamiltonian DLCQ method (or by any other
method) because of the complexities we mentioned above. We will show that
there is a regular and renormalizable Hamiltonian formulation 
that exactly reproduces the results of SDLCQ at large $N_c$, and we will see
that the principal value prescription is a natural consequence of SDLCQ.
Since SDLCQ has been shown to produce finite results in higher dimensions,
these results imply that SDLCQ can be used to generate finite Hamiltonians in
higher dimensions as well. 

We will also find that the SDLCQ Hamiltonian contains a number of irrelevant
operators which we call kronecker terms. These kronecker terms serve to maintain
the supersymmetry at every order of the approximation and act as convergence factors.
If we remove the kronecker terms and calculate the bound-state masses and
wave functions, we find that, while supersymmetry is now broken at every order,
the states very slowly converge to the supersymmetric bound states of SDLCQ. 

We should stress that in the model we study
here  we compactify the null direction $x^-$, and we drop the zero-mode
sector,  which is conventional in DLCQ. We will argue that
dropping some of the zero modes should not be viewed as an omission but 
rather as the
renormalization subtraction that produces the 't Hooft principal value 
prescription
and supersymmetry.

This paper is organized as follows. In section 2 we review some of the
complexities that one finds in theories with dynamical bosons and fermions
even in 1+1 dimensions. In section 3 we review SDLCQ for ${\cal N}=1$ super 
Yang--Mills in 1+1 dimensions, and in section 4 we present the DLCQ Hamiltonian for
this theory that numerically exactly preserves supersymmetry and discuss the
emergence of the principal value prescription and kronecker terms and
our numerical results. We end with some discussion of these results and the
implications for future work in section 5.

\section{ \bf Complexities of DLCQ}

Very few DLCQ calculations involving both dynamical fermions and bosons
have been performed even in 1+1 dimensions because of complexities associated
with renormalization.\footnote{For recent work on Pauli--Villars regularization 
in DLCQ, see Ref.~\cite{PV}.}
We will briefly review these issues. For a more complete
discussion the reader is referred to Ref.~\cite{anp97}.  

The instantaneous Coulomb interactions involving
$2 \rightarrow 2$ parton interactions behave singularly when there is an 
exchange of zero momentum. The same type of Coulomb singularity involving $2
\rightarrow 2$ boson-boson interactions appears in much simpler 
models \cite{dkb94}, and can be shown to cancel a `self-induced' mass term 
(or self-energy) obtained from normal ordering the Hamiltonian. The same
prescription works in the models involving fermions and bosons. There are,
however, finite residual terms after this cancellation is explicitly
performed for the boson-boson and boson-fermion interactions, and they cannot be
absorbed by a redefinition of existing coupling constants.    These residual
terms behave as momentum-dependent mass terms, and the momentum dependence is not
uniquely determined. Examples of these terms can be found in Ref.~\cite{anp97};
they simply multiply the wave functions in the bound-state integral equations.

When one integrates out the left-handed fermion fields, which are dependent
variables (satisfying an equation of the form $\partial_ - \psi_L=F$) in
light-cone quantization, there appears in the light-cone Hamiltonian
a contribution of the form $F^{\dagger}  \frac{1}{{\rm i}\partial_-} F$,
which is singular at $k^+=0$. This singularity is canceled by a (divergent)
momentum-dependent mass term that comes from normal ordering the  
$F^{\dagger}\frac{1}{{\rm i}\partial_-} F$ interactions and performing an
appropriate (infinite) renormalization of the bare boson mass.
The mechanism for cancellation here is different from the Coulombic case,
since it requires specific endpoint relations for wave functions.
For example, the bound-state integral equation governing the behavior
of the two-particle wave function $f(k_1,k_2)$
and the three-particle wave function $h(k_1,k_2,k_3)$ is singular for
vanishing longitudinal momentum fraction in the three-particle wave function.
However, these divergences are  precisely canceled by the momentum-dependent
mass terms. To see this, one must consider the integral equation governing
the three-parton wave function $h(k_1,k_2,k_3)$  at zero momentum fraction.
This leads to the ``ladder relations" 
$h(0,x_2,x_3) \propto  f(x_2, x_3)/\sqrt{x_2} $
and
$ h(x_1,0,x_3) \propto f(x_1,x_3)/\sqrt{x_1} $. The name ``ladder relations"
refers to the fact that they are relations between wave functions with different
numbers of partons. It can then be shown that the singular behavior of the
integral involving the wave function
$h$ can be written in terms of a
momentum-dependent
mass term involving the wave function $f$ by virtue of corresponding
``ladder relations'' \cite{abd97}.
The sum of these divergent contributions exactly
cancels the self-energy contribution. These cancellations generalize to all the
integral equations.

This discussion gives a sense of the difficulties encountered in
setting up DLCQ in non-trivial theories involving both fermions
and bosons even in 1+1 dimensions. What we will see  in
the following sections is that SDLCQ gives a regularization
and renormalization that automatically provides the cancellation
of the self-induced mass and the Coulomb singularity, and there is no need for
the delicate cancellation through ladder relations. In fact, we
will see that every term in the Hamiltonian is finite by itself,
and no conditions or constraints are needed to obtain this
Hamiltonian beyond the SDLCQ formulation. We will see that even
the 't Hooft principal value prescription naturally follows from
SDLCQ.

\section{ \bf Formulation of the bound state problem.}

The light-cone formulation of the
supersymmetric matrix model obtained by dimensionally
reducing ${\cal N} = 1$ $\mbox{SYM}_{2+1}$ to $1+1$ dimensions
has already appeared in \cite{mss95}, to which we refer the
reader for explicit derivations. We simply note here that
the light-cone Hamiltonian $P^-$ is given in terms of the
supercharge $Q^-$ via the supersymmetry
relation $\{Q^-,Q^-\} = 2 \sqrt{2} P^-$, where
\begin{equation}
      Q^-  =  2^{3/4} g \int dx^- \mbox{tr} \left\{
           ({\rm i}[\phi,\partial_- \phi ] + 2 \psi \psi ) \frac{1}{
      \partial_-} \psi \right\}. \label{qminus}
\end{equation}
In the above, $\phi_{ij} = \phi_{ij}(x^+,x^-)$ and
$\psi_{ij} = \psi_{ij}(x^+,x^-)$
are $N_c \times N_c$ Hermitian matrix fields representing the physical
boson and fermion degrees of freedom (respectively) of the theory.
These fields are remnants of the physical transverse degrees of freedom
of the original $2+1$ dimensional theory.
All unphysical degrees of freedom present in the original
Lagrangian have been explicitly eliminated. There are no ghosts.
This is a special feature of light-cone quantization in light-cone
gauge. 

For completeness, we indicate the additional relation
$\{Q^+,Q^+\} = 2 \sqrt{2} P^+$ for the light-cone momentum $P^+$,
where
\begin{equation}
      Q^+  =  2^{1/4} \int dx^- \mbox{tr} \left[
           (\partial_- \phi)^2 + {\rm i}  \psi \partial_- \psi   \right].
\end{equation}
The $(1,1)$ supersymmetry of the model follows from the fact
   $\{Q^+,Q^-\} = 0$.
%
%
%
%
%
%
In order to quantize $\phi$ and $\psi$ on the light cone, we
first introduce the following
expansions at fixed light-cone time $x^{+}=0$:
\begin{eqnarray}
\phi_{ij}(x^-,0)=\frac{1}{\sqrt{2\pi}}\int_0^{\infty}
\frac{dk^+}{\sqrt{2k^+}}\left(a_{ij}(k^+) e^{-ik^+ x^-}+a^\dagger_{ji}(k^+)
e^{ik^+ x^-}\right), \label{phiexp}\\
\psi_{ij}(x^-,0)=\frac{1}{2\sqrt{\pi}}\int_0^{\infty}
dk^+ \left(b_{ij}(k^+) e^{-ik^+ x^-}+b^\dagger_{ji}(k^+)
e^{ik^+ x^-}\right). \label{psiexp}
\end{eqnarray}
We then specify the commutation relations
\begin{equation}
\left[a_{ij}(p^+ ),a^\dagger_{lk}(q^+)\right]=\left\{ b_{ij}(p^+ ),
b^\dagger_{lk}(q^+)\right\}=\delta (p^+ -q^+)\delta_{il}\delta_{jk}
\end{equation}
for the gauge group U($N_c$), or SU($N_c$) in large $N_c$.

For the bound-state eigen-problem
$2P^+ P^- |\Psi> = M^2 |\Psi>$, we may restrict the
subspace of states to those with fixed light-cone momentum $P^+$,
on which $P^+$ is diagonal, and so the bound-state problem is
reduced to the diagonalization
of the light-cone Hamiltonian $P^-$.
Since $P^-$ is proportional to the square of the supercharge $Q^-$,
any eigenstate $|\Psi>$ of $P^-$ with mass squared $M^2$ gives
rise to a natural four-fold degeneracy in the spectrum because
of the supersymmetry algebra---all four states below have the same mass:
\begin{equation}
      |\Psi>, \hspace{4mm} Q^+ |\Psi>,\hspace{4mm}  Q^- |\Psi>,
\hspace{4mm}  Q^+ Q^- |\Psi>.
\end{equation}
Although this four-fold degeneracy is realized in the continuum
formulation of the theory, this property will not
necessarily survive if we choose to discretize the
theory in an arbitrary manner. However, an important
feature of SDLCQ is that it does
preserve the {\it exact} four-fold
degeneracy for any resolution.

The explicit equation for $Q^-$,
in the momentum representation, is obtained by substituting
the quantized field expressions (\ref{phiexp}) and (\ref{psiexp})
directly into the definition of the supercharge (\ref{qminus}).
The result is:
\begin{eqnarray}
\label{Qminus}
Q^-&=& {{\rm i} 2^{-1/4} g \over \sqrt{\pi}}\int_0^\infty dk_1dk_2dk_3
\delta(k_1+k_2-k_3) \left\{ \frac{}{} \right.\nonumber\\
&&{1 \over 2\sqrt{k_1 k_2}} {k_2-k_1 \over k_3}
[a_{ik}^\dagger(k_1) a_{kj}^\dagger(k_2) b_{ij}(k_3)
-b_{ij}^\dagger(k_3)a_{ik}(k_1) a_{kj}(k_2) ]\nonumber\\
&&{1 \over 2\sqrt{k_1 k_3}} {k_1+k_3 \over k_2}
[a_{ik}^\dagger(k_3) a_{kj}(k_1) b_{ij}(k_2)
-a_{ik}^\dagger(k_1) b_{kj}^\dagger(k_2)a_{ij}(k_3) ]\nonumber\\
&&{1 \over 2\sqrt{k_2 k_3}} {k_2+k_3 \over k_1}
[b_{ik}^\dagger(k_1) a_{kj}^\dagger(k_2) a_{ij}(k_3)
-a_{ij}^\dagger(k_3)b_{ik}(k_1) a_{kj}(k_2) ]\nonumber\\
&& ({ 1\over k_1}+{1 \over k_2}-{1\over k_3})
[b_{ik}^\dagger(k_1) b_{kj}^\dagger(k_2) b_{ij}(k_3)
+b_{ij}^\dagger(k_3) b_{ik}(k_1) b_{kj}(k_2)]  \left. \frac{}{}\right\}.
\end{eqnarray}

In order to implement the DLCQ formulation \cite{bpp98} of the
theory, we simply restrict the  momenta $k_1,k_2$ and $k_3$
appearing in the above equation to the following set of allowed
momenta: $\{\frac{P^+}{K},\frac{2P^+}{K},
\frac{3P^+}{K},\dots \}$. Here $K$ is some arbitrary positive
integer and must be sent to infinity if we wish to recover the
continuum  formulation of the theory. The integer $K$  is called the
{\em harmonic resolution}, and $1/K$ measures the coarseness of our
discretization.
Physically, $1/K$ represents the smallest unit of longitudinal
momentum fraction allowed for each parton.  As soon as we implement
the DLCQ procedure, which is  specified unambiguously by the
harmonic resolution $K$, the integrals appearing in the definition
of $Q^-$ are replaced by finite sums, and the eigen-equation is
reduced to a finite matrix problem. In this discrete formulation all operators
containing a zero-momentum operator 
$a_{ij}(0)$,$a_{ij}^\dagger(0)$,$b_{ij}(0)$ or
$b_{ij}^\dagger(0)$ are dropped. We discuss the consequences of this below.

\section{ \bf Hamiltonian Regularization}

In this section we will present the DLCQ Hamiltonian that exactly
reproduces SDLCQ in the large-$N_c$ limit and which is therefore totally
renormalized. We will use a standard operator ordering and suppress all the
indices and variables as follows:

\begin{eqnarray}
a^\dagger a &=& a^\dagger_{ij}(k) a_{ij}(k),\nonumber\\
a^\dagger a^\dagger a a &=&
a^\dagger_{ij}(k_3) a^\dagger_{js}(k_4)a_{ip}(k_1) a_{ps}(k_2)
\delta(k_1+k_2-k_3-k_4),\nonumber\\
a^\dagger a a a &=&
a^\dagger_{ij}(k_4) a_{is}(k_1)a_{sp}(k_2) a_{pj}(k_3)
\delta(k_1+k_2+k_3-k_4),\nonumber\\
a^\dagger a^\dagger a^\dagger a &=&
a^\dagger_{ij}(k_1) a^\dagger_{js}(k_2)a^\dagger_{sp}(k_3) a_{ip}(k_4)
\delta(k_1+k_2+k_3-k_4).
\end{eqnarray}

There is a well defined and unambiguous method to find the DLCQ
Hamiltonian. We start from the discrete SDLCQ supercharge, square it, and
then normal order the results. In the continuum formulation this is, of course,
a trivial restatement that the Hamiltonian is the square of the supercharge;
however, it is not a trivial statement in the discrete formulation
since the zero-mode operators have been dropped. In the normal ordering process
one contracts various operators to form the Hamiltonian operator, but in
SDLCQ the zero-mode operators are missing, and, therefore, the Hamiltonian will
be missing operators that would have been formed from the contraction of the
zero modes. In addition, of course, the zero modes that are normally dropped
in DLCQ are also dropped here.  After considerable algebra one arrives at the
normal ordered form of the square of the discrete supercharge, which is our
renormalized DLCQ Hamiltonian
\begin{eqnarray}
P^-&=&\frac{g^2N_c}{4\pi}\int_0^\infty dk_1 \frac {\mu^2(k_1)}{k_1}
(a^\dagger a +b^\dagger b)+\frac{g^2}{4\pi}\int_0^\infty dk_1dk_2dk_3dk_4[
\nonumber\\ &+&
   A_1 b^\dagger b^\dagger b b+
   A_2 (b^\dagger b b b - b^\dagger b^\dagger b^\dagger b) +
   B_1 a^\dagger a^\dagger a a +
	B_2 (a^\dagger a a a + a^\dagger a^\dagger a^\dagger a)
\nonumber\\ & + &
	C_1 b^\dagger b^\dagger a a + C_2 a^\dagger a^\dagger b b +
   C_3 b^\dagger a^\dagger b a + C_4 a^\dagger b^\dagger a b +
   C_5 b^\dagger a^\dagger a b + C_6 a^\dagger b^\dagger b a
   \nonumber \\ &+&
   D_1 (a^\dagger a b b - a^\dagger b^\dagger b^\dagger a) +
   D_2 (a^\dagger b a b - b^\dagger a^\dagger b^\dagger a) +
   D_3 (a^\dagger b b a - b^\dagger b^\dagger a^\dagger a)
\nonumber\\ &+&
   D_4 (b^\dagger b a a + b^\dagger a^\dagger a^\dagger b) +
   D_5 (b^\dagger a b a + a^\dagger b^\dagger a^\dagger b) +
   D_6 (b^\dagger a a b + a^\dagger a^\dagger b^\dagger b)],
   \end{eqnarray}
where 
\begin{equation}
\mu^2(k_1)=\int_0^{k_1} dk_2\frac{{(k_1+k_2)}^2}{k_2{(k_1-k_2)}^2},
\end{equation}
and the other coefficients are given by
\begin{eqnarray}
A_1 &=& PV\frac{2}{(k_4-k_2)^2}-\frac{2}{(k_1+k_2)^2} -
\delta_{1,3}(\frac{2}{k_1^2}+\frac{2}{k_2^2}),
\nonumber\\
A_2 &=& \frac{2}{(k_2+k_3)^2}-\frac{2}{(k_1+k_2)^2},
\nonumber\\
B_1 &=& \frac{1}{\sqrt{4k_1k_2k_3k_4}}
\left(\frac{(k_1-k_2)(k_3-k_4)}{(k_1+k_2)^2}-
PV\frac{(k_1+k_3)(k_2+k_4)}{(k_4-k_2)^2}\right),
\nonumber\\
B_2 &=& \frac{1}{\sqrt{4k_1k_2k_3k_4}}
\left(\frac{(k_3-k_2)(k_1+k_4)}{(k_3+k_2)^2}+
\frac{(k_1-k_2)(k_3+k_4)}{(k_1+k_2)^2}\right),
\nonumber\\
C_1&=& \frac{1}{\sqrt{k_1k_2}}
\left(\frac{k_1-k_2}{(k_1+k_2)^2} +PV\frac{1}{2(k_3-k_1)}
-\frac{(k_1-k_2)\delta_{1,3}}{k_3k_4}\right),
\nonumber\\
C_2&=& \frac{1}{\sqrt{k_3k_4}}
\left(\frac{k_4-k_3}{(k_1+k_2)^2} +PV\frac{1}{2(k_3-k_1)}
-\frac{(k_4-k_3)\delta_{1,3}}{k_1k_2}\right),
\nonumber\\
C_3&=& -\frac{1}{\sqrt{k_2k_4}}
\left(PV\frac{k_2+k_4}{(k_2-k_4)^2} -\frac{1}{2(k_1+k_2)}
-\frac{2k_4\delta_{1,3}}{k_1^2}\right),
\nonumber\\
C_4&=& -\frac{1}{\sqrt{k_1k_3}}
\left(PV\frac{k_1+k_3}{(k_1-k_3)^2} -\frac{1}{2(k_1+k_2)}
-\frac{2k_1\delta_{1,3}}{k_2^2}\right),
\nonumber\\
C_5&=& -\frac{1}{\sqrt{4k_1k_4}}
\left(PV\frac{1}{(k_3-k_1)} +\frac{1}{(k_1+k_2)}
+\frac{2(k_1+k_2)\delta_{1,3}}{k_1k_2}\right),
\nonumber\\
C_6&=& -\frac{1}{\sqrt{4k_2k_3}}
\left(-PV\frac{1}{(k_3-k_1)} +\frac{1}{(k_1+k_2)}
+\frac{2(k_1+k_2)\delta_{1,3}}{k_2k_3}\right),
\nonumber\\
D_1&=& \frac{1}{\sqrt{k_1k_4}}
\left(\frac{k_1+k_4}{(k_2+k_3)^2} -\frac{1}{2(k_1+k_2)}\right),
\nonumber\\
D_2&=& -\frac{1}{\sqrt{4k_2k_4}}
\left(\frac{1}{(k_2+k_3)} -\frac{1}{(k_1+k_2)}\right),
\nonumber\\
D_3&=& -\frac{1}{\sqrt{k_3k_4}}
\left(\frac{k_3+k_4}{(k_1+k_2)^2} -\frac{1}{2(k_2+k_3)}\right),
\nonumber\\
D_4&=& \frac{1}{\sqrt{k_2k_3}}
\left(\frac{k_3-k_2}{(k_2+k_3)^2} +\frac{1}{2(k_1+k_2)}\right),
\nonumber\\
D_5&=& -\frac{1}{\sqrt{4k_1k_3}}
\left(\frac{1}{(k_2+k_3)} +\frac{1}{(k_1+k_2)}\right),
\nonumber\\
D_6&=& -\frac{1}{\sqrt{k_1k_2}}
\left(\frac{k_2-k_1}{(k_1+k_2)^2} -\frac{1}{2(k_2+k_3)}\right).
\nonumber\\
\end{eqnarray}
It is understood that the integrals become finite sums for the
DLCQ calculation. 
The forms of the counterterms are included in the
coefficients, $A_i, B_i, C_i,$ and $D_i$. 

In comparing this result to SDLCQ, the
first obvious feature is that the Hamiltonian has many more terms than the
supercharge. Numerically this is quite significant since each of these terms
has to act on the entire fock space to calculate the Hamiltonian matrix, and,
therefore, the DLCQ approach is more time consuming. In
SDLCQ the supercharge must be squared before it is used to calculate
the spectrum; however, squaring a numerical matrix can be done very
efficiently.

The second thing to notice is the appearance of the principal value
regularization of the singularities in terms $A_1, B_1$ and all the $C_i$ terms.
The use of the principal value is a common feature in DLCQ dating back to the
't Hooft model \cite{tho}. {\it What is new here is that, if we trace the origin of
these subtractions back to the  SDLCQ formulation, we see that they occur
because we dropped the zero-mode  operators.} The operators that are discarded
by the principal value prescription are just the set of operators that appear
when normal ordering the zero modes. In the discrete calculation we
include theta functions\footnote{$\theta(x) = 1$ for $ x > 0$ and zero
otherwise.} to enforce the missing zero-mode contributions.  We always
find two identical terms  that are missing the zero-mode
contribution, but one comes with a 
$\theta(k_3-k_1)$ and while the other comes with $\theta(k_1-k_3)$. In the
continuum limit these combine to one, but in the discrete calculation they
combine to give the principal value. While normally one considers dropping
any mode in a calculation an undesirable approximation, here we see that {\it
dropping zero modes is equivalent to a renormalization subtraction } and, in fact,
an unexpectedly good subtraction.

Some time ago another method of treating this singularity, which produced a 
different numerical result, was suggested by Wu \cite{wu} and by Mandelstam 
and Leibbrandt \cite{mal}. It is only recently that the connection between 
this subtraction and the principal value 't Hooft prescription was fully 
understood. It was shown in \cite{bas00} that, in fact, the 't Hooft 
prescription is equivalent to an infinite set of topological terms in the 
other prescriptions. The fact that subtracting the intermediate zero modes 
in SDLCQ automatically leads one to this correct result is clearly one of 
the attractive features of this method.

Another important feature of this Hamiltonian, that one would not see  in the
usual DLCQ Hamiltonian, are the terms with the kronecker delta.\footnote{We  are
thinking of the integrals as sums when we write these terms as kronecker deltas.}
These terms, which we call kronecker terms, are zero in the continuum
formulation and scale to zero as the  resolution gets large in the DLCQ
formulation.     These terms arise in the discrete calculation from terms of the
form $1-(\theta(k_1-k_3) +\theta(k_3-k_1))$. 
They destroy two particles and replace them with two
particles with the same momentum.  The coefficients include a momentum dependent 
factor that scales to zero as the resolution goes to infinity.\footnote{For
earlier work on this type of zero-mode contribution, see Ref.~\cite{Wivoda}.}

To fully understand the significance of these kronecker terms, it
is helpful to write one of them in discrete form. Without any leading numerical
factors the $A_1$ term in the Hamiltonian has the following discrete form:
\begin{eqnarray}
P^-&\propto& Lg^2\sum_{{n_1,n_2,n_3,n_4=1},{n_4\neq n_2}}^\infty B^\dagger(n_3)
B^\dagger(n_4) B(n_1) B(n_2) {2 \over (n_4 -n_2)^2} \delta_{n_1+n_2-n_3-n_4}
\nonumber \\
&-& Lg^2\sum_{n_1,n_2,n_3,n_4=1}^\infty B^\dagger(n_3)
B^\dagger(n_4) B(n_1) B(n_2) {2 \over (n_1 +n_2)^2} \delta_{n_1+n_2-n_3-n_4}
\nonumber \\
&-& Lg^2\sum_{n_1,n_2=1}^\infty B^\dagger(n_1)
B^\dagger(n_2) B(n_1) B(n_2)( {2 \over (n_1)^2} +{2 \over (n_2)^2}).
\end{eqnarray}
In this form only $g$ and $L$ carry dimensions.
Now let us go to infinite resolution and convert the sums to
integrals. The detailed translation between the discrete and continuous
formulation is given in Ref.~\cite{mss95}. For our purposes it is sufficient to
note that 
\[
k_i={n_i
\pi \over L}, \quad \quad P^+={K \pi \over L}, \quad \quad 
{\pi \over L}\sum_{n=1}^\infty \rightarrow \int_0^\infty, \quad
\quad B(n)\rightarrow
\sqrt{{\pi \over L}}b(n).
\]
Then  the contribution of this term to $P^-$  translates to
\begin{eqnarray}
P^- &\propto& PV\int_0^\infty \prod_{i=1}^4dk_i \delta(k_1+k_2-k_3-k_4)
b^\dagger (k_3)b^\dagger (k_4)b(k_1)b(k_4) 
(\frac{2}{(k_4-k_2)^2}-\frac{2}{(k_1+k_2)^2}) \nonumber \\
&&+ 
{P^+ \over K}\int_0^\infty dk_1 dk_2 
b^\dagger (k_1)b^\dagger (k_2)b(k_1)b(k_2) 
(\frac{2}{(k_1)^2}+\frac{2}{(k_2)^2}). 
\end{eqnarray}
The appearance of $1/K$ in the second term shows that this term is irrelevant.
As $K$ goes to infinity this term goes away.
%
\begin{figure}
\begin{tabular}{cc}
\psfig{file=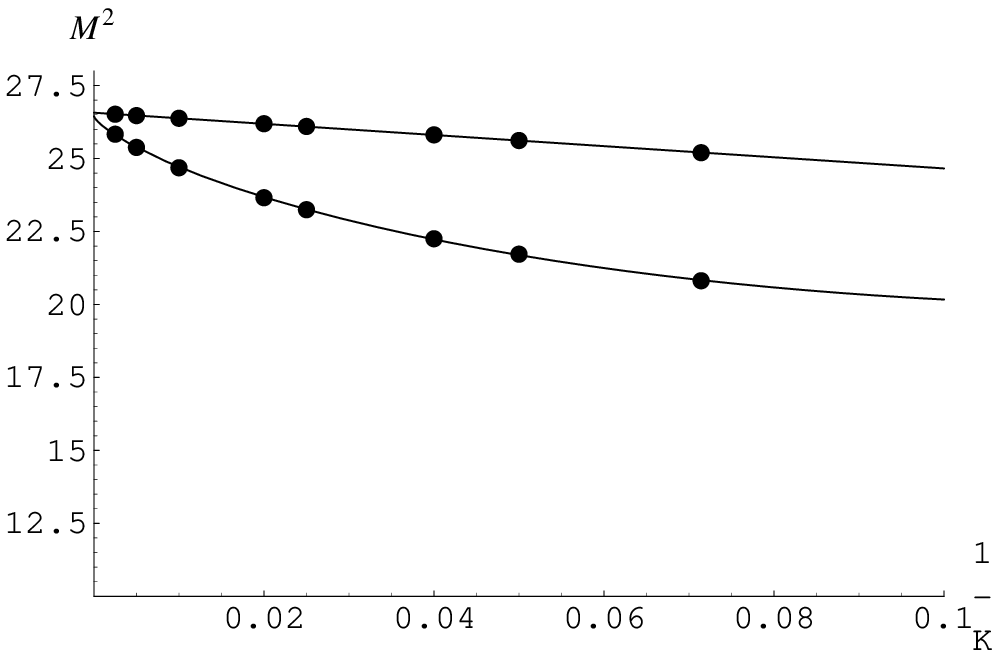,width=3.2in}  &
\psfig{file=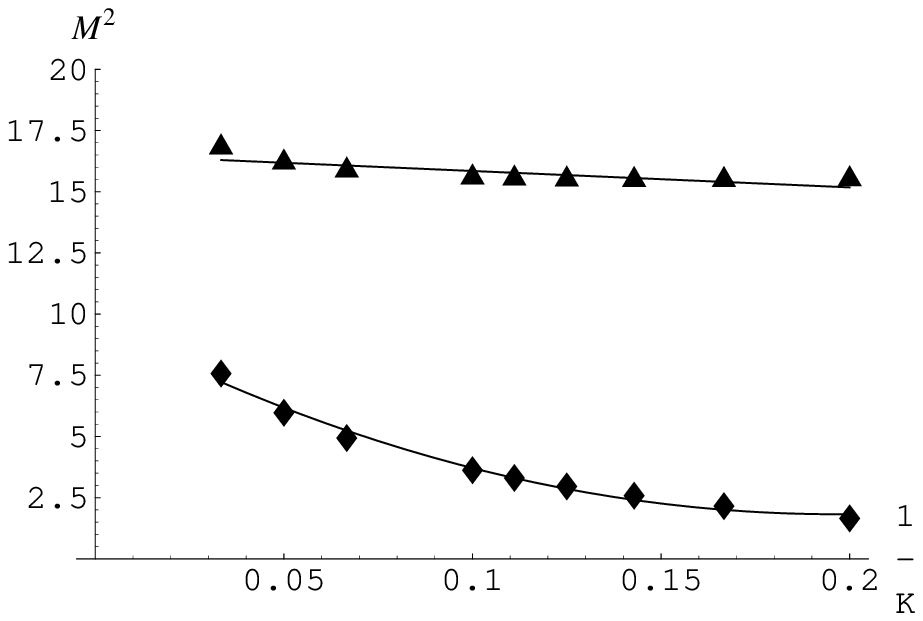,width=3.2in}  \\
(a) & (b) 
\end{tabular}
\caption{ ${\pi M^2\over g^2 N_c}$ vs. ${1 \over K}$, where $K$ is the
resolution, for (a) the Kutasov model and (b) full ${\cal N}=1$ SYM.
In each case the upper curve includes the kronecker terms, but the
lower one does not.
\label{convergence}}
\end{figure}
%

We first saw this type of term in the SDLCQ formulation \cite{alp99b} of a
supersymmetric model proposed by Kutasov \cite{Kutasov93}. The Kutasov model is
essentially the above Hamiltonian with an appropriately chosen mass term.
The kronecker terms serve two functions; they serve to enforce the
supersymmetry at every resolution and act as  convergence factors. Dropping
these terms, one gets the same numerical results at infinite resolution, but the 
convergence is very slow, and supersymmetry is broken at finite resolution. 
In addition, massless states are driven to negative $M^2$ values.  In
Fig.~\ref{convergence}a we show our results for the Kutasov model \cite{alp99b}. 
The top flat
curve is SDLCQ while the lower curve is the result when the kronecker terms are
removed. In this calculation we had to go to very high resolution to actually
see the two different versions of the theory converge \cite{alp98}. We see very
similar results in Fig.~\ref{convergence}b for the full SYM theory. We find
that the effect of the kronecker terms  is very large and the convergence is
very slow without these terms. We cannot go to high enough resolution to
obtain the complete convergence seen in the Kutasov model. 

To get to these very high resolutions in either model we had to truncate the
fock space \cite{alp99b}.  To get to the highest resolution in the full SYM theory
we truncated the basis to five particles. We have compared the effect of this
truncation up to resolution $K=10$ and find it leads to only small changes in the
mass of the bound state. There are many bound states in the region of this bound
state that we do not show. To make sure we are referring to the same state in the
two calculations  we take the inner product of the wave functions. We find that
wave functions in the two calculations have better than a
$90 \%$ overlap. This also gives us information about the effect on the wave 
functions of dropping the kronecker terms.

This renormalized Hamiltonian is a starting point to begin the investigation of
non-supersymmetric theories that cannot be written as the square of  a
supercharge. In addition there already exist SDLCQ calculations in 2+1
dimensions \cite{alp99a,hhlp99} which can be used to produce non-perturbatively
renormalized Hamiltonians in 2+1 dimensions. 

Finally, we should remind the reader that there is a set of zero modes that we
have not addressed here, the diagonal zero modes of $A^+$ and its
superpartners. These modes are discussed elsewhere \cite{alpt98}.  They give
rise to modes that wind around the compact space and  to the $N_c$ degenerate
vacua of this model.


\section{Conclusions}

The non-perturbative renormalization of a light-cone quantized Hamiltonian gauge
theory with dynamical bosons and fermions can be a complicated  and ambiguous
procedure even in 1+1 dimensions.  As a result there have only been a few DLCQ
calculations of this type. A promising approach appears to be the very natural 
marriage of DLCQ with supersymmetry. Together they
generate a powerful numerical technique, SDLCQ, for solving exactly
supersymmetric theories. To date many exactly supersymmetric  theories in
1+1 \cite{lup99} and 2+1 \cite{alp99a,hhlp99,hpt01} dimensions have been solved
using SDLCQ, and the results of these calculations  have been used to address  a
number of fundamental issues in string theory and related areas.

In this paper we revisited DLCQ and found the DLCQ Hamiltonian that exactly
preserve supersymmetry. We present a procedure for producing non-perturbative
renormalized DLCQ Hamiltonians that are free from the complexities that one
normally encounters in DLCQ. We found a unique set of counterterms in DLCQ that
achieve this  result and that have a number of important properties.
Surprisingly we discovered that dropping zero modes in SDLCQ should be viewed as
the renormalization subtraction that produces the 't Hooft principal value
prescription in DLCQ. This is particularly appealing since the principal value
prescription has recently been shown \cite{bas00} to automatically include a
series of topological corrections not included in other 
prescriptions \cite{wu,mal}.

In addition we find a set of irrelevant terms, which we call kronecker terms,
that scale away at infinite resolution. They make the Hamiltonian
exactly  supersymmetric at every resolution and serve as convergence factors. The
importance of numerical convergence factors should not be overlooked; they can
be the difference between a successful calculation and one that has to await
larger and faster computers.  We presented results for ${\cal N}=1$ SYM models with 
and without the kronecker terms and found that they have a large effect and that the
convergence without kronecker terms is very slow. When we include the
kronecker terms, the large-$N_c$ SDLCQ results for the Kutasov model match
those of Ref.~\cite{alp99b}. 

This improved  technology represents a first step toward extending the
advantages of SDLCQ to DLCQ and treating models with supersymmetry
breaking. There already exist SDLCQ calculations in 2+1
dimensions \cite{alp99a,hhlp99,hpt01} which can be used to produce
non-perturbatively renormalized Hamiltonians in 2+1 dimensions, and we hope to
use this new renormalization technique to study theories that break
supersymmetry.

\section{Acknowledgments}
This work was supported in part by the US Department of Energy. One of the authors
(S.P) would like to acknowledge the hospitality of the Aspen Center of Physics
where part of the work was completed. The authors would like to acknowledge
conversations with U. Trittmann and O. Lunin.

\end{document}